\documentclass[12pt]{article}
\usepackage{epsfig}
\usepackage{amssymb}
\usepackage{aas_macros}

\newcommand{\Lam}{\Lambda}

\newcommand{\mpc}{{\rm Mpc}}

\newcommand{\kms}{\rm km\,s^{-1}}

\newcommand{\Msun}{M_{\odot}}

\newcommand{\hinv}{h^{-1}}
\newcommand{\himpc}{\hinv{\rm\,Mpc}}

\newcommand{\himsun}{\hinv{\Msun}}

\newcommand{\Om}{\Omega_{\rm m}}
\newcommand{\Ol}{\Omega_{\Lam}}
\newcommand{\Ob}{\Omega_{\rm b}}

\newcommand{\Lbox}{{L_{\rm box}}}

\newcommand{\Mstar}{M_{\star}}

\newcommand{\beq}{\begin{eqnarray}}
\newcommand{\eeq}{\end{eqnarray}}

\begin{document}

\begin{center}
{\bf \Large{Star Formation History of Dwarf Galaxies in Cosmological Hydrodynamic Simulations}}\\
\vspace{0.2cm}
{\it Kentaro Nagamine}\\
Department of Physics and Astronomy, University of Nevada, Las Vegas\\
4505 S. Maryland Pkwy, Box 454002, Las Vegas, NV 89154-4002, U.S.A.\\
Email: kn@physics.unlv.edu
\end{center}


\vspace{0.2cm}
\begin{center}
{\bf \large Abstract}
\end{center}
\vspace{-0.3cm}
We examine the past and current work on the star formation 
(SF) histories of dwarf galaxies in cosmological hydrodynamic simulations.  
The results obtained from different numerical methods are still somewhat 
mixed, but the differences are understandable if we consider the numerical 
and resolution effects. 
It remains a challenge to simulate the episodic nature of SF history
in dwarf galaxies at late times within the cosmological context of 
a cold dark matter model.  More work is needed to solve the mysteries of 
SF history of dwarf galaxies employing large-scale hydrodynamic simulations 
on the next generation of supercomputers.


\section{Introduction}
\label{sec:intro}

Dwarf galaxies play unique roles in the cosmological studies of galaxy 
formation.  Low-mass galaxies contribute the most to the cosmic star 
formation rate (SFR) density at low redshift, and in contrast 
the star formation in high-mass galaxies seems to cease at $z\gtrsim 1$ 
\cite{Juneau05, LeFloch05}; the so-called `downsizing' effect in 
galaxy formation.  
This global trend of star formation is presumably driven 
by the feedback effects by supernovae (SNe) and supermassive black holes,  
as well as the cosmological effects such as the expansion and the 
reionization of the Universe.  All of these effects can suppress the 
star formation, eventually giving rise to the characteristic shape of 
galaxy luminosity function.  
The question that is not fully answered yet is, ``Which physical 
processes have the strongest impact, and which galaxies are affected 
the most?'' 
Studying the star formation in dwarf galaxies gives us 
useful clues not only on the physics of star formation, but also on the 
feedback processes that cause the downsizing in galaxy formation and 
how the galaxy luminosity function has been shaped over time. 

Observations of stellar populations suggest that the star formation in 
dwarf galaxies \cite{Dolphin03, Dolphin05, McQuinn09, Searle73, Weisz08} 
is sporadic, 
separated by millions to billions of years, even in isolated systems. 
What causes the episodic SF history?
Past interaction with other galaxies or merger events are obvious possibilities
to explain the SF activity in these systems, but the lack of tidal debris
in the outer regions of dwarf galaxies argues against such scenarios 
\cite{Dellen08}.  Then the remaining likely possibility is the instability 
in the local interstellar medium (ISM), however, many interesting questions
remain.  For example, why is it {\it now} that they are undergoing 
active star formation?  What determines the epoch and the duration of 
star formation?
Is it the local physics or the cosmological processes that are more 
important in determining the downsizing effect?

While the recent episodic SF activity is observed in dwarf galaxies, 
most dwarf galaxies in the Local Group are also dominated by the old stellar 
populations (ages of $\gtrsim$10\,Gyr) with an occasional mixture of 
younger stars \cite{Gnedin00d, Grebel97, Mateo98, vanBergh99}.  
This suggests that the main formation epoch of those dwarf galaxies
were before $z\sim 6$. 

It is important to address the above questions using the {\it ab initio} 
cosmological hydrodynamic simulations, in which the gas dynamics is 
simulated self-consistently from early universe to the present time.
Although there have been much numerical work on the formation of dwarf
galaxies \cite{Gnedin00a, Gnedin06, Hoeft06, Kravtsov04a, Mashchenko08, 
Ricotti05, Tassis08, Tassis03}, 
usually the work is presented in the context of reionization or 
the missing satellite problem, and most publications do not 
present the SF histories of dwarf galaxies in their simulations.  

As we describe in the next section, explaining the episodic nature of 
SF history in low-mass systems remains a challenge for many reasons 
in the framework of cold dark matter (CDM) model. 
The purpose of this short article is to record what we know
on the SF history of dwarf galaxies in the past and current cosmological 
hydrodynamic simulations.


\section{Cosmological Hydrodynamic Simulations}

\subsection{Star Formation Model}

In order to simulate the formation and evolution of dwarf galaxies, 
one has to model the collapse of gas clouds and subsequent star formation.  
In most cosmological simulations, 
the star formation is modeled by creating a star particle from a 
parcel of gas (either the gas in an Eulerian cell or a gas particle 
in the case of smoothed particle hydrodynamics [SPH]) when the following
criteria are satisfied \cite{Cen92, Katz92}:
(i) the region is overdense ($\delta > \delta_c$);
(ii) converging gas flow ($\nabla \cdot {\bf v} < 0$); 
(iii) cooling rapidly ($t_{cool} < t_{dyn}$); 
and (iv) Jeans unstable ($m_{gas} > m_{Jeans}$).
One would expect that if the condition (iii) or (iv) is satisfied, 
the other conditions are also likely to be met, but more rigorous analyses 
would be required to assess which criterion is the most important one. 
There are other variants of SF models, but the discussion will be deferred
to another place \cite{Choi09b}.

Simulations of galaxy formation require vast dynamic ranges in both 
mass and space: from molecular clouds ($\sim$parsecs) to groups/clusters of 
galaxies ($\sim$mega-parsecs).  With the currently available computational 
resources, 
it is still difficult to resolve the details of molecular clouds while 
simultaneously simulating the formation of thousands of galaxies on the 
scales of $\gtrsim$10\,Mpc.
In other words, we can reliably identify the sites of galaxy formation
on large-scales and simulate the average SFR of each galaxy, but the 
small-scale instabilities of multiphase ISM on sub-parsec scales caused 
by, e.g., turbulence, cannot be followed in detail in cosmological 
simulations.  As we will describe in the next section in more detail, 
the earlier Eulerian cosmological simulation \cite{Nag01b} was not 
able to reproduce
the episodic nature of SF history, even though it properly simulated
the gas dynamics on large scales ($\gtrsim$20\,kpc).  
This suggests that the sporadic star formation in dwarf galaxies is 
driven by the instabilities on small scales, which is difficult to 
simulate properly with the current resolution limits. 

The above limitation is a typical criticism directed towards cosmological 
simulations, however, the cosmological simulations can 
calculate the amount of gas that fall into the region correctly. 
Combined with a star formation law, we can calculate the overall SFR of 
the system, which is valid within the limitations of the model and still 
is relevant to the subject of this article. 
We simulate the gas dynamics at intermediate scales ($\sim$kpc) as 
accurately as possible, and treat the small-scale physics 
with a sub-grid/particle model for star formation and SN feedback based on 
our astrophysical knowledge. In this approach, the SFR is calculated as
$\dot{\rho}_\star = c_\star (\rho_{gas}/t_\star)$, where $c_\star$ is 
the SF efficiency and $t_\star$ is the SF time-scale.  Both of these 
parameters are usually adjusted so that the simulation result matches 
the observed Kennicutt-Schmidt law \cite{Kennicutt98, Schmidt59}.  
One can further adopt a sub-grid/particle model \cite{Springel03b, Yepes97} 
to calculate the cold gas density $\rho_{cold}$, and use this in place of 
$\rho_{gas}$ in the above SF law. The projected SFR generally follows the 
Kennicutt law well \cite{Choi09a, Nag04f, Schaye08, Springel03b}, but 
the consequences of the adopted SF model (e.g., the SF threshold density 
or the equation of state) must be studied further \cite{Choi09b}.


\subsection{Past and Current Work}
  
Ref.~\cite{Nag01b} showed that, using an Eulerian cosmological hydrodynamic 
simulation, the star formation ceased at high redshift in the dwarf galaxies 
that survived to $z=0$.  In their simulation, only 7\% of stars formed 
below $z=1$ in galaxies with stellar masses 
$2\times 10^8 < M_\star(z=0) < 2\times 10^9 \himsun$, and no stars formed 
at $z<1$ in galaxies with $M_\star(z=0) < 2\times 10^8 \himsun$. 
This result is consistent with the general expectation in the CDM model, 
in which the small systems form early on and the accretion of material 
stops thereafter, as the residual gas is easily swept out by the 
SN feedback or evaporated by photoionization \cite{Dekel86, Efstathiou92, 
Gnedin00a, Navarro97, Quinn96, Weinberg97}. 
The conclusion of Ref.~\cite{Nag01b} was that, in a CDM universe, dwarf
galaxies found today are dominated by old stars; they consist predominantly
of stars 10 Gyrs old and do not show recent SF activity. 

However, as we outlined in Section~\ref{sec:intro}, the observed SF histories 
of dwarf galaxies exhibit sporadic SF activity at late times, and we
wonder whether the result of Ref.~\cite{Nag01b} was affected by the 
resolution limitation.  Eulerian mesh simulations lose 
the physical spatial resolution at late times owing to the cosmic 
expansion. On the other hand, they generally have a larger mesh number 
compared to 
the number of particles employed in cosmological SPH simulations, hence 
have higher baryonic mass resolution at early times than SPH simulations.  
(If the number of dark matter particles and the box sizes are comparable, 
then the resolution in the initial gravitational field would be similar 
in the two methods.) 
This numerical trend could explain the efficient conversion of gas into 
stars at early times and the lack of SF activity at late times in the 
Eulerian mesh simulation of Ref.~\cite{Nag01b}, because the simulation 
may underestimate the cooling rate of gas when it cannot resolve the 
the density fluctuation of gas on scales below the physical mesh size, 
as they approach to the present time.   
A related notable advantage of the Eulerian mesh simulation over the 
SPH method is that it is better at modeling the gas in low density regions, 
therefore it is more suitable for the study of Ly-$\alpha$ forest 
\cite{Cen94}. 

\begin{figure}[t]
\begin{center}
\resizebox{8.1cm}{!}{\includegraphics{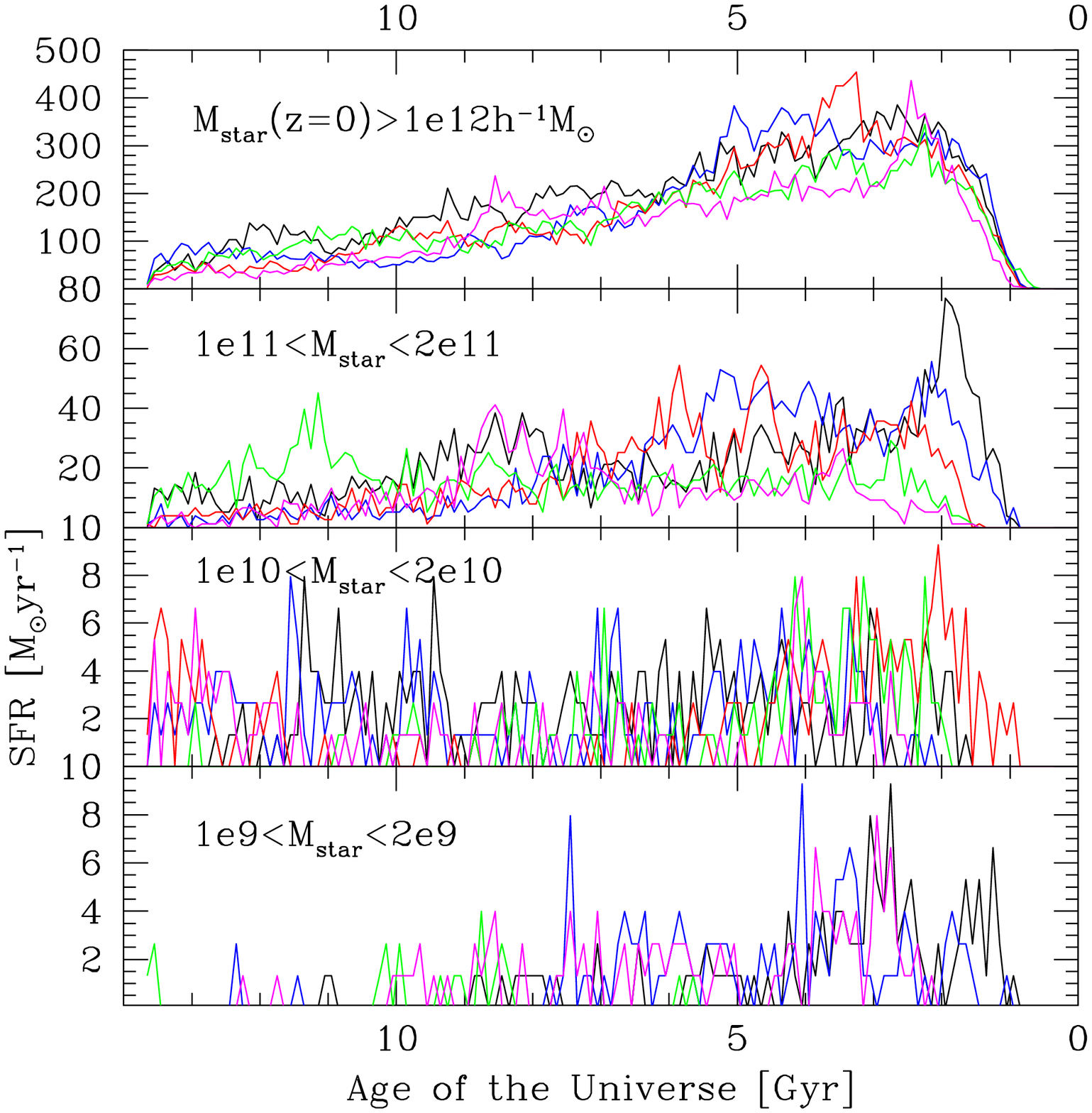}}
\resizebox{8.1cm}{!}{\includegraphics{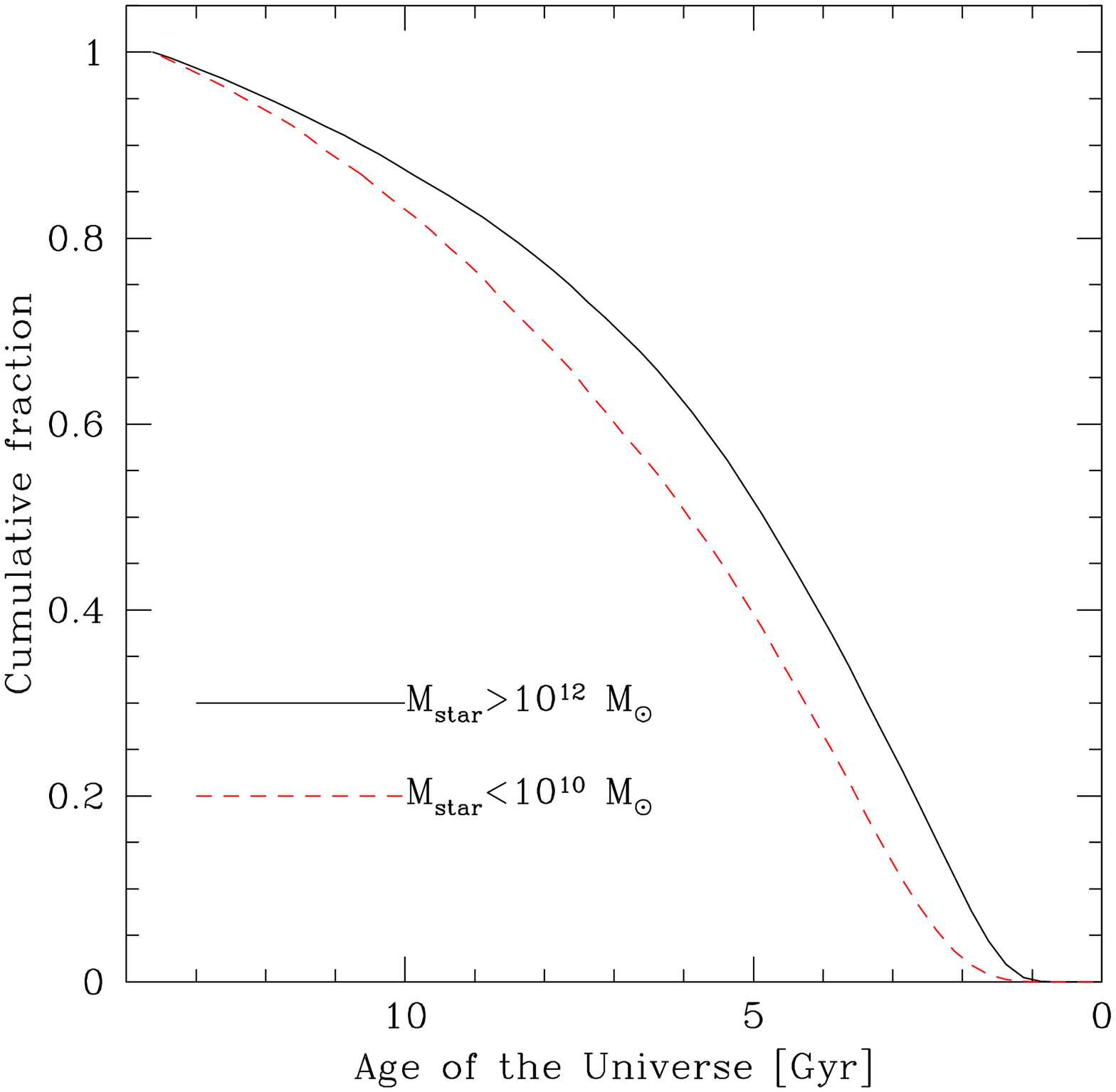}}
\caption{\small {\it Left:} 
We compare some examples of star formation histories with 10$^8$\,yr bins 
in a cosmological SPH simulation of a comoving box size $\Lbox = 100\himpc$ 
and $2\times 400^3$ (gas and dark matter) particles.  In each panel, five 
different galaxies in the indicated stellar mass ranges (in units of 
$\himsun$) at $z=0$ are shown with different colors.  
Stars formed in all the progenitors are co-added for each galaxy. 
{\it Right:} Cumulative stellar mass fraction as a function of cosmic time
for all the galaxies with $\Mstar(z=0) > 10^{12}\himsun$ and $\Mstar(z=0) < 10^{10}\himsun$. 
}
\label{fig:sfhist}
\end{center}
\end{figure}

In contrast to the Eulerian mesh simulation, the SPH simulation is 
more suitable to simulate the late-time evolution of high-density 
regions owing to its Lagrangian nature. Therefore one might expect
that the SPH simulations perform better in modeling the sporadic SF history
at late times in dwarf galaxies. 
In the left panel of Figure~\ref{fig:sfhist}, we compare some examples of 
SF history in a 
cosmological SPH simulation of a comoving box size $\Lbox = 100\himpc$ and 
$2\times 400^3$ (gas and dark matter) particles.  The adopted cosmological 
parameters are consistent with the latest WMAP5 results \cite{Komatsu09}: 
$(\Om,\Ol,\Ob,\sigma_8, h)= (0.26, 0.74, 0.044, 0.80, 0.72)$, where 
$h=H_0 /(100\kms\,\mpc^{-1})$.  The masses of dark matter, gas and star 
particles are $(m_{dm}, m_{gas}, m_{star}) = (9.4\times 10^8, 1.9\times 10^8, 
8.5\times 10^7)\,\himsun$.    
In each panel, we indicate the range of stellar masses of the selected 
galaxies at $z=0$. 
Each SF history includes all the stars formed in the progenitors of 
the current galaxy. The SF histories of the most massive galaxies with 
$M_\star > 10^{12}\himsun$ peak at $t=2-3$\,Gyr, and gradually decline
thereafter in a roughly exponential manner, consistently with the results of 
Ref.~\cite{Nag01b}.  In the bottom panel, the SF histories of dwarf galaxies 
with $1\times 10^9 < M_\star < 2\times 10^9\himsun$ are shown. In this 
simulation, the dwarf galaxies continue to form stars sporadically at 
late times, even at $t>10$\,Gyrs. 

In the right panel of Figure~\ref{fig:sfhist}, we compare the cumulative
stellar mass fraction that formed in galaxies with 
$\Mstar(z=0) > 10^{12}\himsun$ and $\Mstar(z=0) < 10^{10}\himsun$
as a function of the age of the universe.  This figure shows that, 
in a relative sense, the more massive galaxies form their stars earlier 
than the lower mass galaxies, in qualitative agreement with the observational 
trend (e.g., \cite{Thomas05}) of downsizing.

Even though our new SPH simulations \cite{Choi09b} seem to be more 
successful in modeling the qualitative trend of star formation in different 
galaxies, we may still have difficulties in reproducing the correct 
number density of ``red \& dead'' massive galaxies at $z=1-2$ or the 
ultraluminous infrared galaxies (ULIRGs), as examined by 
Refs.~\cite{Nag05a, Nag05d}. Interestingly, their work showed that the 
Eulerian simulation exhibited more sporadic SF history for the massive 
galaxies than the SPH simulation at intermediate redshifts of $z=1-3$.  
The difference in the nature of SF history was perhaps due to a combination 
of differences in the SF models and the effectiveness of feedback, 
as well as the numerical resolution reached in the different simulations.

\section{Discussion \& Conclusion}

As described in the previous section, the results from different numerical 
methods are still somewhat mixed, but we understand their qualitative 
differences when we consider the resolution and numerical effects. 
The main factor seems to be the numerical resolution, and one way to get
around this problem might be to utilize the adaptive mesh refinement 
(AMR) technique (e.g., \cite{Tassis03}).  However, as we pointed out in 
Ref.~\cite{Oshea05}, current AMR codes require substantially larger 
computational resources to obtain an equivalent result on the dark matter
halo mass function as the SPH codes do at early times.  Therefore it is 
difficult to simulate a large sample of dwarf galaxies in a large 
cosmological volume, and the simulation is usually not run down to $z=0$ 
\cite{Gnedin09, Tassis03}.  

Perhaps the best way to view the current situation is that each code
captures the essence of dwarf galaxy formation at different epochs.
The Eulerian mesh codes can capture the very early starbursts in low-mass 
halos better than the SPH codes, whereas the SPH codes can capture the 
gas infall and star formation in dwarf galaxies at late times owing to 
their higher spatial resolution in high-density regions.  
Both of these processes (i.e., early and late star formation) probably 
took place in the real Universe as Ref.~\cite{Ricotti05} proposed: 
some dwarf galaxies being the true fossils of the pre-reionization era, 
some being dominated by the late star formation at low redshift, 
and the rest being the mixture of the two. 

Our hope is that, within the next decade, the results from different 
numerical techniques (Eulerian mesh, SPH, and AMR) will converge, and 
provide a consistent picture of dwarf galaxy formation.  Then we will 
have a much better idea on the physical processes that shaped the galaxy 
luminosity function and how the downsizing effect of galaxy formation 
is caused.  To achieve this goal, we still have to overcome a huge 
dynamic range from subparsec to Mpc scales, and it will require a 
peta-scale supercomputer of next generation, such as {\it Blue\,Waters} 
\cite{Bluewaters}.


\bigskip
\leftline{\bf \large Acknowledgment}
\medskip

\noindent
The author is grateful to Jun-Hwan Choi who kindly allowed me to 
present the results of our cosmological SPH simulations that we are 
collaborating on. The full results of our work will be presented 
elsewhere \cite{Choi09b}.
This work is supported in part by the National Science Foundation 
grant AST-0807491, the National Aeronautics and Space
Administration under Grant/Cooperative Agreement No. NNX08AE57A issued
by the Nevada NASA EPSCoR program, the National Science Foundation 
through TeraGrid resources provided by the Texas Advanced Computing 
Center and the San Diego Supercomputer Center, and the President's 
Infrastructure Award at UNLV.



\footnotesize

\begin{thebibliography}{}
\bibitem{Cen92} Cen, R. \& Ostriker, J. P., 1992, \apj, 399, L113
\bibitem{Cen94} Cen, R. \& Ostriker, J. P., 1994, \apj, 437, L9
\bibitem{Choi09a} Choi, J.-H. \& Nagamine, K., 2009, \mnras, 393, 1595
\bibitem{Choi09b} Choi, J.-H. \& Nagamine, K., 2009, \mnras, submitted (arXiv:0909.5425)
\bibitem{Dekel86} Dekel, A. \& Silk, J., 1986, \apj, 303, 39
\bibitem{Dellen08} Dellenbusch, K. E., Gallagher, J. S., Knezek, P. M., et al. 2008, AJ, 135, 326
\bibitem{Dolphin03} Dolphin, A. E., Saha, A., Skillman, E., et al. 2003, AJ, 126, 187
\bibitem{Dolphin05} Dolphin, A. E., Weisz, D. R., Skillman, E., et al. 2005, arXiv:astro-ph/0506430
\bibitem{Efstathiou92} Efstathiou, G., 1992, \mnras, 256, 43
\bibitem{Gnedin00a} Gnedin, N. Y., 2000, \apj, 542, 535 
\bibitem{Gnedin00d} Gnedin, N. Y., 2000, \apj, 535, L75
\bibitem{Gnedin06} Gnedin, N. Y., 2006, \apj, 645, 1054
\bibitem{Gnedin09} Gnedin, N. Y., Tassis, K., \& Kravtsov, A. V., 2009, \apj, 697, 55
\bibitem{Grebel97} Grebel, E. K., 1997 Rev. Mod. Astron., 10, 29 
\bibitem{Hoeft06} Hoeft, M., Yepes, G., Gottl\"{o}eber, S., et al. 2006, \mnras, 371, 401
\bibitem{Juneau05} Juneau, S., Grazebrook, K., Crampton, D., et al. 2005, ApJL, 619, L135
\bibitem{Katz92} Katz, N., 1992, \apj, 391, 502
\bibitem{Kennicutt98} Kennicutt, R. C. Jr., 1998, \apj, 498, 541
\bibitem{Komatsu09} Komatsu, E., Dunkley, J., Nolta, M. R., et al., 2009, ApJS, 180, 330
\bibitem{Kravtsov04a} Kravtsov, A. V., Gnedin, N. Y., \& Klypin, A. A., 2004, \apj, 609, 482
\bibitem{LeFloch05} Le Floc'h, E., Papovich, C., Dole, H., et al. 2005, \apj, 632, 169
\bibitem{Mashchenko08} Mashchenko, S., Couchman, H. M. P., \& Sills, A., 2005, \apj, 624, 726
\bibitem{Mateo98} Mateo, M., 1998, \araa, 36, 435
\bibitem{McQuinn09} McQuinn, K. B. W., Skillman, E. D., Cannon, J. M., et al. 2009, ApJ, 695, 561
\bibitem{Nag01b} Nagamine, K., Fukugita, M., Cen, R. et al. 2001, \apj, 558, 497
\bibitem{Nag04f} Nagamine, K., Springel, V., \& Hernquist, L., 2004, \mnras, 348, 435 
\bibitem{Nag05a} Nagamine, K., Cen, R., Hernquist, L., et al. 2005, \apj, 627, 608
\bibitem{Nag05d} Nagamine, K., Cen, R., Hernquist, L., et al. 2005, \apj, 618, 623
\bibitem{Navarro97} Navarro, J. F. \& Steinmetz, M. 1997, \apj, 478, 13
\bibitem{Oshea05} O'Shea, B. W., Nagamine, K., Springel, V., et al. 2005, ApJS, 160, 1 
\bibitem{Quinn96} Quinn, T., Katz, N., \& Efstathiou, G. 1996, MNRAS, 278, L48
\bibitem{Ricotti05} Ricotti, M. \& Gnedin, N. Y., 2005, \apj, 629, 259
\bibitem{Rocha00} Rocha-Pinto, H. J., Scalo, J., Maciel, W. J., et al., 2000. ApJL, 531, L115
\bibitem{Schaye08} Schaye, J. \& Dalla Vecchia, C., 2008, \mnras, 383, 1210
\bibitem{Schmidt59} Schmidt, M., 1959, \apj, 129, 243
\bibitem{Searle73} Searle, L., Sargent, W. L. W., Bagnuolo, W. G., 1973, \apj, 179, 427 
\bibitem{Springel03b} Springel, V. \& Hernquist, L., 2003, \mnras, 339, 289 
\bibitem{Tassis08} Tassis, K., Kravtsov, A. V., \& Gnedin, N. Y., 2008, \apj, 672, 888
\bibitem{Tassis03} Tassis, K., Abel, T., Bryan, G. L., et al., 2003, \apj, 587, 13
\bibitem{Thomas05} Thomas, D., Maraston, C., Bender, R., et al. 2005, \apj, 621, 673
\bibitem{vanBergh99} van den Bergh, S., 1999, A\&A Rev., 9 273
\bibitem{Weinberg97} Weinberg, D. H., Hernquist, L., \& Katz, N. 1997, \apj, 477, 8
\bibitem{Weisz08} Weisz, D. R., Skillman, E. D., Cannon, J. M., et al., 2008, \apj, 689, 160
\bibitem{Yepes97} Yepes, G., Kates, R., Kohklov, A., et al., 1997, \mnras, 284, 235
\bibitem{Bluewaters} http://www.ncsa.uiuc.edu/BlueWaters/
\end{thebibliography}

\end{document}